\documentclass[12pt,titlepage,natbib209,twocolumn]{prop}

\usepackage[letterpaper,margin=1in]{geometry}
\usepackage{html}
\usepackage{graphicx,deluxetable}
\usepackage[z-times,expert,alpha]{tmsfnts}
\usepackage{amssymb}

\def\fnl{f_{\rm NL}}
\def\vk{\vec{k}}
\begin{document}
\title{Non-Gaussianity from large-scale structure surveys}
\author{Licia Verde \thanks{email:{\tt liciaverde@icc.ub.edu}}}
\date{{\it\small ICREA (Instituci\'o Catalana de Recerca i Estudis Avan\c cat) \&\\  ICC-UB (Instiuto de Ciencias del Cosmos, Universitat de Barcelona,\\ Marti i Franques 1, 08028 Barcelona, ES\\[3mm]
Paper submitted to the special issue ``Testing the Gaussianity and Statistical Isotropy of the Universe'' of Advances in Astronomy}\\[5mm]
\begin{minipage}[h]{6.5in}
{\bf Abstract.} 
\end{minipage}
 With the advent of  galaxy surveys which provide large samples of galaxies or galaxy clusters over a volume comparable to the horizon size (SDSS-III, HETDEX, Euclid, JDEM, LSST, Pan-STARRS, CIP etc.) or mass-selected large cluster  samples over a large fraction of the extra-galactic sky (Planck, SPT, ACT, CMBPol, B-Pol), it is timely to investigate what constraints these surveys can impose on primordial non-Gaussianity. I illustrate here three different approaches: higher-order correlations of the three dimensional galaxy distribution,  abundance of rare objects (extrema of the density distribution), and the large-scale clustering of  halos (peaks of the density distribution). Each of these avenues has its own advantages, but, more importantly, these approaches are highly complementary under many respects.}
\maketitle

\section{Introduction}
\label{sec:intro}

The recent advances in  the understanding of the origin and evolution of the Universe  have been driven by the advent of high-quality data, in unprecedented amount (just think of WMAP and SDSS for example). Despite this, most of the information about cosmological parameters come from the analysis of  a massive compression of the data: the power spectrum of their statistical fluctuations over the mean. The power spectrum is a complete statistical description of a random field only if it is Gaussian.  

Even the simplest inflationary models predict deviations from  Gaussian initial conditions. These deviations are expected to be small, although ``small" in some models may be ``detectable". For a thorough review of inflationary non-Gaussianity see \cite{BKMR04}, for our purpose it will be sufficient to say that
to describe inflation-motivated departures from gaussian initial conditions many write (\cite{Salopekbond90, Ganguietal94, VWHK00, KS01}:
\begin{equation}
\Phi=\phi+f_{NL}\left( \phi^2-\langle \phi^2\rangle \right)\,.
\label{eq:fnl}
\end{equation}
 
Here $\phi$ denotes a gaussian field and  $\Phi$ denotes 
Bardeen's gauge-invariant potential, which, 
on sub-Hubble scales reduces to the usual Newtonian peculiar 
gravitational potential, up to a minus sign.  In the literature, there are two conventions for Eq.~(\ref{eq:fnl}): the large-scale structure (LSS) and the Cosmic Microwave Background (CMB) one. In the LSS  convention  $\Phi$ is linearly extrapolated at $z=0$; in the CMB convention $\Phi$ is instead primordial:  thus $f^{LSS}_{\rm NL}=g(z=\infty)/g(0) f_{\rm NL}^{CMB}\sim 1.3 f^{CMB}_{NL}$, where $g(z)$ denotes the linear growth suppression factor relative to an Einstein-de-Sitter Universes. In the past few years it has become customary to always report $ f_{\rm NL}^{CMB}$ values even if, for simplicity as it will be clear below, one carries out the calculations  with $f_{\rm NL}^{LSS}$.

 While for simplicity one may just assume $f_{\rm NL}$ in eq. \ref{eq:fnl} to be a constant (yielding the so-called {\it local} model or {\it local}-type)  in reality the expression is more complicated and $f_{NL}$ is scale and configuration dependent. 
In general, the non-Gaussianity  is specified by writing down the bispectrum of $\Phi$.  For example, one can see that for the local model the bispectrum is:
\begin{equation}
B_{\Phi}(k_1,k_2,k_3)=2 f_{\rm NL} P_{\phi}(k_1) P_{\phi}(k_2) +{\rm  2\,  cyc.}
\end{equation}
Where $P$ denotes the power spectrum and it is often assumed that $P_{\phi}=P_{\Phi}$;  ``cyc." denotes two cyclic terms over $k_1, k_2, k_3$.

It has been shown \cite{BabichCreminelliZaldarriaga} that  for non-Gaussianity of the local type the bispectrum is dominated by the so-called {\it squeezed} configurations, triangles where one wavevector length is much smaller than the other two. Models such as the curvaton for example have a non-Gaussianity of the local type. 
 Standard, single-field slow roll inflation also yields a local non-Gaussianity but with an unmeasurably small $\fnl$ (see \cite{BKMR04} and references therein). On the other hand, many inflationary models have an {\it equilateral}-type non Gaussianity, i.e. the bispectrum is dominated by  equilateral triangles \cite{BabichCreminelliZaldarriaga}. Ref. \cite{creminelli1, creminelli2}  have proposed a functional form which closely approximates the 
behavior of the inflationary bispectrum  and which is useful for efficient data-analysis:
\begin{eqnarray}
B(\vk_1,\vk_2,\vk_3)&=& 6\fnl\left\{-P(k_1)P(k_2)+2 {\rm cyc.} \right.\nonumber \\
&-&2[P(k_1)P(k_2)P(k_3)]^{2/3}\\ 
&+& \!\!\!\!\left.P^{1/3}(k_1)P^{2/3}(k_2)P(k_3) +5 {\rm cyc.}\right\}. \nonumber
\end{eqnarray}
Note that the same numerical value for $\fnl$ gives rise to a larger skewness in the local case than in the equilateral case (e.g. see \cite{Loverdeetal07}), explaining why the  CMB constraints on $f_{\rm NL}$ are weaker for the equilateral case \cite{creminelli1,creminelli2, komatsuwmap08}.

Specific deviations from a single field, slow roll, canonical kinetic energy, Bunch-Davies vacuum, leave their specific signature on the  bispectrum ``shape" (i.e. the dependence of $B$ on the shape of the  triangle made by the three $\vk$ vectors), see discussion in  \cite{komatsuwhitepaper} and references therein. 

Non-Gaussianity therefore offers a probe of aspects of inflation (namely the interactions of the inflaton) that are difficult to probe by other means (i.e., measuring the shape of the primordial power spectrum and properties of the stochastic background  of gravity waves). So, how could primordial non-Gaussianity be tested?

 One could look at the early Universe: by looking at CMB anisotropies we can probe cosmic fluctuations 
at a time when their statistical distribution should have been close 
to their original form but the signal is small. On the other hand, one could analyze the statistics of the large-scale structures, close to the present-day, when the signal is larger, but   this is a more complicated approach, since gravitational  instability  (for the dark matter distribution) and bias   (for galaxies or clusters of galaxies) introduce non-Gaussian features in an initially Gaussian field and they mask the signal one is after.
Finally, the abundance of rare events (such as galaxy clusters and high-reshift galaxies) probes the tails of the PDF of the density field, which are extremely sensitives to deviations from Gaussianity.
Here, I will concentrate on the signature of non-Gaussianity on  large-scale structure (i.e. at redshift $z \lesssim 1$) as they can be traced by  galaxy surveys (i.e., I will not consider wide field weak gravitational lensing surveys); other contributions to this review will focus on non-Gaussian  signatures on the  CMB,  thus here it will be sufficient  to give only a very  brief  and succinct introduction to the subject.

At recombination the density fluctuations are small( $\delta_{\Phi} \sim 10^{-5}$), and the CMB temperature fluctuations are directly related to $\Phi$ making this a very clean probe.  However, effectively only one redshift can be tested giving us only a  2-dimensional  information. 

The most widespread technique for testing Gaussianity in the CMB is to use the CMB bispectrum:
\begin{equation}
\langle a_{\ell_1}^{m_1} a_{\ell_2}^{m_2} a_{\ell_3}^{m_3}\rangle=B_{\ell_1\ell_2\ell_3} \left(^{\ell_1 \,\,\,\,\ell_2 \,\,\,\, \ell_3}_{m_1\,\, m_2\,\, m_3}\right) 
\end{equation}
where the $a_{\ell}^m$ are the coefficients of the spherical harmonic  expansion  of the CMB temperature fluctuation: $\Delta T/T=\sum_{\ell m} a_{\ell}^m Y_{\ell}^m$ and the presence of the 3-J symbol  ensures that the bispectrum is defined if $l_1+l_2+l_3={\rm even}$,  $\ell_j+\ell_k \ge l_i \ge |l_j-\ell_k|$ (triangle rule) and that $m_1+m_2+m_3=0$.
It should be however clear that secondary CMB anisotropies and foregrounds also induce a CMB bispectrum which can mask or partially mimic the signal  see e.g., Refs. \cite{Goldberg&Spergel-II, KS01, VerdeSpergel01, CoorayHu2001, MangilliVerde09, Serra-2008, Hansonetal09} and references therein.

In the last few years, this area of research has received an impulse, motivated by the recent full sky CMB data from WMAP. In particular  it has been shown that the constraints can be greatly improved by effectively  ``reconstructing" the potential $\Phi$ from  CMB temperature and polarization data rather than simply using the temperature bispectrum alone \cite{YadavWandelt05,YadavKomatsuWandelt07}. This technique would yield constraints  on non-Gaussianity of the  local type of $\Delta f_{\rm NL} \sim 1 $ for an ideal experiment and $\Delta f_{\rm NL} \sim 3$ for the Planck satellite. This is particularly promising as $f_{\rm NL} $ of order unity or larger is produced by broad classes of  inflationary models (see e.g., \cite{BKMR04} and  references therein).

Currently, the most stringent constraints for the local type  are $27 <f_{\rm NL} < 147$ at the 95\% confidence (central value  87) from WMAP 3 yr data \cite{YadavWandelt08}; and from the WMAP 5 years data, $-9< f_{\rm NL} < 111$  at the 95\% confidence level (central value 55)  \cite{komatsuwmap08} and  $-4 < \fnl < 80$ \cite{Smith09}. Despite the heated debate on wether $f_{\rm NL}=0$ is ruled out or not, the two measurement are not necessarily in conflict:  the two central values differ by only about 1$\sigma$; different, although not independent, data sets were used with different galactic cuts, and the maximum multipole considered in the analyses is also different. 
What  makes the subject very interesting, is that, if the central value for $f_{NL}$ is truly around $60$, forthcoming data will yield a highly-significant detection.

\section{Higher-order correlations}
\label{sec:hoc}
Theoretical considerations  (see  discussion in e.g., \cite{komatsuwhitepaper} and references therein) lead us to define primordial non-Gaussianity by its bispectrum.  While in principle there may be types of non-Gaussianity  which would be more directly related to higher-order correlations (e.g., \cite{gnlpaper} and references therein), and  while a full description of a non-Gaussian distribution would require the specification of all the  higher-order correlations, it is clear that quantities such as the bispectrum  enclose information about the phase correlation between $k$-modes. In the Gaussian  case, different Fourier mode are uncorrelated (by definition of Gaussian random phases) and a statistic like the power spectrum  does not carry information about phases.  
The bispectrum is the lowest-order correlation with  zero expectation value in a Gaussian random field. But, even if the initial conditions were Gaussian,   non-linear evolution due to gravitational instability generates a non-zero bispectrum.  In particular, gravitational instability has its own ``signature" bispectrum, at least in the next-to-leading order in cosmological perturbation theory \cite{Catelanetal95}:
\begin{equation}
B(\vk_1,\vk_2,\vk_3)=2P(k_1),P(k_2)J(\vk_1,\vk_2)+ 2 {\rm cyc.}
\end{equation}
where   $J(\vk_1,\vk_2)$ is the gravitational instability ``kernel" which depends very weakly on cosmology and for an Einstein-de-Sitter Universe is:
\begin{equation}
J(\vk_1,\vk_2)=\frac{5}{7}+\frac{\vk_1\cdot \vk_2}{2k_1k_2}\left(\frac{k_1}{k_2}+\frac{k_2}{k_1}\right)+\frac{2}{7}\left(\frac{\vk_1\cdot \vk_2}{k_1 k_2}\right)^2\,.
\end{equation}
In the highly non-linear regime the  detailed form of the kernel changes, but it is something that could be computed and calibrated  by extending perturbation theory beyond the next-to-leading order  and by comparing with numerical N-body simulations (see other contributions in this issue). 
It was recognized a decade ago \cite{VWHK00} that this signal is quite large compared to any expected primordial non-Gaussianity and that the primordial signal ``redshifts away"  compared to the gravitational signal. In fact,  a  primordial signal given by a local type of non-Gaussianity parameterized by a given $\fnl$,  would affect the late-time dark matter density bispectrum with a  contribution of the form
\begin{equation}
B^{\fnl\, local}(\vk_1,\vk_2\vk_3,z)=
\end{equation}
$$2\fnl P(k_1)P(k_2)\frac{{\cal F}(\vk_1,\vk_2)}{D(z)/D(z=0)}+ 2 {\rm cyc}.$$
where $D(z)$ is the linear growth function which in an Einstein-de Sitter universe goes like $(1+z)^{-1}$ and 
\begin{equation}
{\cal F}= \frac{{\cal M}(k_3)}{{\cal M}(k_1){\cal M}(k_2)}\,; \,\,\,{\cal M}(k)=\frac{2}{3}\frac{k^2 T(k)}{H_0^2\Omega_{m,0}}\,,
\end{equation}
$T(k)$ denoting the transfer function,  $H_0$ the Hubble parameter and $\Omega_{m,0}$ the matter density parameter.
Clearly the two contributions have different scale and redshift dependence and the two kernel shapes in configuration space are different, thus,  making the two component, at least in principle and for high signal-to-noise, separable.

Unfortunately, with galaxy surveys, one does not observe the dark matter distribution directly. Dark matter halos are believed to be hosts for galaxy formation,  and  different galaxies  at different redshifts populate halos following different prescriptions.
In large-scale structure studies, often the assumption of linear, scale independent bias is made. A linear bias will not introduce a non-zero bispectrum in a Gaussian field and  its effect on a field with a non-zero  bispectrum is only to re-scale its bispectrum  amplitude.    This is, however, an approximation,  possibly  roughly valid at large scales for dark matter halos,  and when looking at the power spectrum,  but unlikely to be true in detail. 
To go beyond the linear bias assumption, often the assumption of quadratic bias  is made, where the relation between dark matter overdensity field and galaxy field is specified by two parameters: $b_1$ and $b_2$: $\delta_g(x)=b_1\delta_{\rm DM}(x)+b_2(\delta_{\rm DM}^2-\langle \delta_{\rm DM}^2 \rangle)$; $b_1$ and $b_2$ are assumed to be scale-independent (although this assumption must break down at some point) but they can vary with redshift. Clearly,  a quadratic bias will introduce non-Gaussianity  even on an initially Gaussian field. In summary, for local non-Gaussianity and scale-independent quadratic bias we have \cite{MVH97,VWHK00}:
\begin{equation}
B(\vk_1,\vk_2,\vk_3,z)=2 P(k_1)P(k_2) b_1(z)^3\times 
\end{equation}
$$ \left[ \fnl \frac{{\cal F}(\vk_1,\vk_2)}{D(z)} + J(\vk_1,\vk_2) +\frac{b_2(z)}{2 b_1(z)}\right]+ cyc.\\ $$
 Before the above expression can be compared to observations it needs to be further complicated by redshift space distortions (and shot noise). Realistic surveys  use the redshift as a proxy for distance, but gravitationally-induced peculiar velocities distort  the redshift-space galaxy  distribution. We will not go into these details here as including redshift space distortions (and shot noise) will not change    the gist of the message.
 
 From a practical point of view, it is important to note that photometric surveys, although in general can cover larger volumes that spectroscopic ones, are not suited for this analysis: the projection effects due to the photo-z smearing along the line-of-sight is expected to suppress significantly the sensitivity of the measured bispectrum to the shape of the primordial one (see e.g., \cite{VHM00,LSSTBook}).
 
Ref. \cite{VWHK00} concluded that ``CMB is likely to provide a better probe of such [local] non-Gaussianity". Much more recently, \cite{SefusattiKomatsu07}  revisited the issue and  found that, assuming  a given -known-  redshift dependence of the  ($b_1, b_2$) bias parameters and  an all sky  survey  from $z=0$ to $z=5$ with  a galaxy number density of at least $5 \times 10^{-4} h^3/$Mpc$^3$, the galaxy bispectrum can provide  constraints on  the $\fnl$ parameter  competitive with CMB.  However, for all planned surveys,  the forecasted errors are much larger than Planck forecasted errors.
This holds qualitatively also for the equilateral case. 

While the gravitationally-induced non-Gaussian signal in the bispectrum has been detected  to high statistical significance (see \cite{Verde2df} and references therein; other contributions to this issue);  the non-linear bias signature is not uncontroversial, and there have been so far no detection of any  extra (primordial) bispectrum contributions.

Of course one could also consider higher-order correlations. One of the advantages of considering e.g., the trispectrum is that, contrary to the bispectrum, it has very weak non-linear growth \cite{Verdetrispc}, but has the disadvantage that the signal is de-localized: the number of possible configurations grows fast with the dimensionality $n$ of the $n$-point function!

In summary, higher-order correlations as observed in the CMB or in the evolved Universe, can be used to determine the bispectrum {\it shape}. The two approaches should be seen as complementary  as they are affected by different systematic effects and probe different scales. The next two probes we consider have a less rich sensitivity to the bispectrum shape, but their own peculiar advantages. 
 
\section{The Mass Function}
\label{sec:MF}
The abundance of collapsed objects (dark matter halos as traced e.g.,  by galaxies and galaxy clusters)
contains important information about the properties of initial conditions on galaxy and clusters scales.
The Gaussian assumption plays a central role in analytical predictions for the abundance 
and statistical properties of the first objects to collapse in the Universe. In this context, the 
formalism proposed by Press \& Schechter \cite{PS74}, with its later extensions and 
improvements,  has 
become the Ôstandard loreÕ for predicting the number of collapsed dark matter halos as a function 
of redshift. However, even a small deviation from Gaussianity can have a deep impact on those 
statistics which probe the tails of the distribution. This is indeed the case for the abundance of 
high-redshift objects like galaxies and clusters  at $z \gtrsim 1$ 
 which correspond to high peaks, 
i.e. rare events, in the underlying dark matter density field. Therefore, even small deviations from 
Gaussianity might be potentially detectable by looking at the statistics of high-redshift systems.
Before proceeding let us introduce some definitions.

We are interested in predictions for rare objects, that is the collapsed objects that form in 
extreme peaks of the density field $\delta(x)=\delta\rho/\rho$. The statistics of collapsed objects can be 
described by the statistics of the density perturbation smoothed on some length scale $R$
(or equivalently a mass scale $M = 4/3 \pi R^3\rho$), $\delta_R$.

To incorporate non-Gaussian initial conditions into predictions for the smoothed density field, we need an expression for the probability distribution function (PDF) for $\delta_R$. For 
a particular real-space expansion like Eq. (\ref{eq:fnl}), one may make a formal change of variable in 
the Gaussian PDF to generate a normalized distribution \cite{MVJ00}.  However, this may not be possible in general and  the change of variables does not work for  the smoothed cumulants of the 
density field. In general, the PDF for a generic non-Gaussian distribution  can be written exactly as a function of the cumulant's generating function ${\cal W}_R$ for the smoothed density field:
\begin{equation}
{\cal P}(\delta_R)d\delta_R=\int\frac{d\lambda}{2\pi}\exp[-i\lambda \delta_R+{\cal W}_R(\lambda)]d\delta_R
\label{eq:PDFexact}
\end{equation}
with
\begin{equation}
{\cal W}_R(\lambda)=\sum _{n=2}^{\infty}\frac{(i\lambda)^n}{n!}\mu_{n,R}
\end{equation}
where $\mu_{n,R}$ denote the cumulants and for example $\mu_{2,R}=\sigma_R^2=\langle \delta_R^2\rangle$ and
 the skewness $\mu_{3,R}$ is related to  the normalized skewness of the  smoothed density field $S_{3,R}=\mu_{3,R}/\mu_{2,R}^2$. It is useful to  define a ``skewness per $\fnl$ unit" $S_{3,R}^{\fnl=1}$ so that  $S_{3,R}=\fnl S_{3,R}^{\fnl=1}$.
 The skewness $\mu_{3,R}$ is related to the underlying bispectrum by:
 \begin{equation}
 \mu_{3,R}\!=\!\int \frac{d^3\! k_1 d^3k_2 d^3k_3}{(2\pi^7)}   B_{\delta,R}(\vk_1,\vk_2,\vk_3)   \delta^D_{\vk_1+\vk_2+\vk_3}
 \end{equation}
where  $ \delta^D$ denotes the Dirac delta function and  $B_{\delta,R}$ denotes the bispectrum of the $\delta$ overdensity field smoothed on scale $R$. It is related to the potential one trivially by remembering the  Poisson equation: $\delta_R(\vk)={\cal M}(k)W_R(k)\Phi(\vk)$. Here, $W_R(k)$ denotes the smoothing kernel:  usually taken to be the Fourier transform of the top hat window.
In any practical application the dimensionality of the integration can be reduced by collapsing the expressing $\vk_3$ as a function of $\vk_1$ and $\vk_2$.

It is important at this point to make a small digression to specify definitions of key quantities. Even in linear theory, the normalized skewness of the density field depends on redshift; however in the Press \& Schechter  framework one should always use linearly extrapolated quantities at $z=0$. In this context therefore, when writing $S_{3,R}=\fnl S_{3,R}^{\fnl=1}$, if $S_{3,R}^{\fnl=1}$ is that if the density field extrapolated linearly at $z=0$ then $\fnl $ must be the LSS one and not the CMB one.

 \begin{figure}[t]
\centerline{\includegraphics[width=1.0\linewidth]{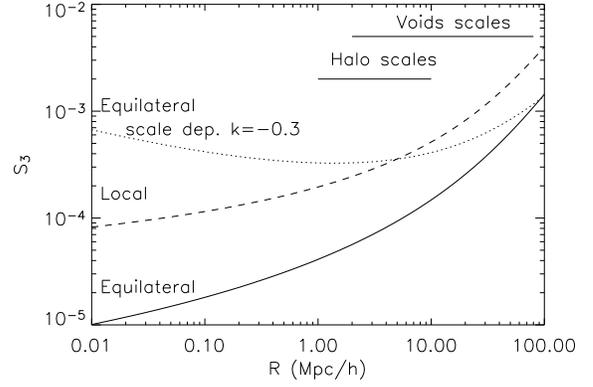}}
\caption{Skewness $S_{3,R}$ of the density field at $z=0$ as a function of the smoothing scale $R$ for different types of non-Gaussianity. Figure reproduced  from \cite{KVJ09}.}
\label{fig:ske}
\end{figure}
To compute  the  abundance of collapsed objects fro the PDF one will then follow the  Press \& Schechter 
swindle: first compute 
\begin{equation}
{\cal P}(>\delta_c|z,R)=\int_{\delta_c(z)}^{\infty}d\delta_R {\cal P}(\delta_R)
\end{equation}
 where $\delta_c$ denotes the critical threshold for collapse; then  the number of collapsed objects is:
 \begin{equation}
 n(M,z) dM=2 \frac{3 H_0^2\Omega_m}{8 \pi G M}\left| \frac{d{\cal P}(>\delta_c|z,R)}{dM}\right|\,.
\label{eq:massfnPS}
 \end{equation}
 Note that the redshift dependence is usually enclosed only in $\delta_c$: $\sigma_M$ is computed on the field linearly extrapolated at $z=0$, and $\delta_c(z)=\Delta_c(z)D(z=0)/D(z)$ and $\Delta_c(z)$ depends very weakly on redshift and $\Delta_c(z=0)\simeq 1.68$ .
 
 Eq. (\ref{eq:massfnPS}) however cannot be computed analytically and exactly starting from Eq. (\ref{eq:PDFexact}): some approximations need to be done in order to obtain an analytically manageable expression. Two approaches have been taken so far in the literature which we will briefly  review below.
 
\subsection{MVJ approach.}

The authors of Ref.\cite{MVJ00} proceed by first performing the integration over $\delta_R$ to obtain an exact expression  for ${\cal P}(>\delta_c|z,M)$. At this point they expand the generating functional  to the desired order, e.g., keeping  only terms up to the skewness, then perform a Wick rotation to change variables and  finally a saddle-point approximation to evaluate the remaining integral. The saddle point approximation is very good for large thresholds $\delta_c/\sigma_M>>1$, thus for rare and massive peaks.
For the final expression for the mass function they obtain:
$$\hat{n}(M,z)=2\frac{3H_0^2\Omega_{m,0}}{8\pi GM^2}\frac{1}{\sqrt{2\pi}\sigma_M} 
\exp\left[-\frac{\delta_*^2}{2\sigma_M^2}\right] \times $$
\begin{equation}
\left| \frac{1}{6}\frac{\delta_c^2}{\sqrt{1-S_{3,M}\delta_c/3}} 
\frac{dS_{3,M}}{d\ln M} +\frac{\delta_*}{\sigma_M}
\frac{d \sigma_M}{d\ln M}\right|  \nonumber 
\label{eq: massfnMVJ}
\end{equation}
where $\sigma_M$ denotes the {\it rms} value of the density field, the 
subscript $M$ denotes that the density field has been smoothed on a scale 
$R(M)$ corresponding to $R(M)=[M 3/(4\rho)]^{1/3}$, and 
$\delta_*=\delta_c\sqrt{1-\delta_c S_{3,M}/3}$. 

This derivation shows that  the mass function in principle depends on all cumulants, but that if non-Gaussianity is described by a bispectrum (and all higher order connected correlations are assumed to be zero or at least negligible),  it depends only on the skewness. The mass function  does not carry explict information about the shape of non-Gaussianity. Nevetheless  for a given numerical value of $\fnl$ the skewness can have different amplitude and scale dependence for different models of non-Gaussianity, as illustrated in Fig. (\ref{fig:ske}).

\subsection{LMSV approach}

The authors of Ref. \cite{Loverdeetal07} (LMSV) instead proceed by using the saddle point approximation in the expression for ${\cal P}(\delta_R)$ and then  using the Edgeworth expansion truncated at the desired order. The resulting simplified expression for the PDF can then be integrated to obtain ${\cal P}(>\delta_c|z,R)$ and derived to obtain the mass function.
The final mass function in this approximation is given by:
$$
\hat{n}(M,z)=2\frac{3H_0^2\Omega_{m,0}}{8\pi GM^2}\frac{1}{\sqrt{2\pi}\sigma_M} 
\exp\left[-\frac{\delta_c^2}{2\sigma_M^2}\right] \times 
$$
$$
 \left[  \frac{d \ln \sigma_M}{dM}\left(\frac{\delta_c}{\sigma_M}+\right. 
\frac{S_{3,M}\sigma_M}{6} \left( \frac{\delta_c^4}{\sigma_M^4}
-2\frac{\delta_c^2}{\sigma_M^2}-1\right) \right)   
$$
\begin{equation}
\left. +\frac{1}{6}\frac{dS_{3,M}}{dM}
\sigma_M\left(\frac{\delta_c^2}{\sigma_M^2}-1\right)  \right] \,.
\label{eq:massfnloverde}
\end{equation}
Without knowledge of all higher cumulants one is forced to use approximate expressions for the mass function.
By  truncating  the Edgeworth expansion at the linear order in the  skewness, the resulting PDF is no longer positive-definite. The number of terms that should be kept in the expansion depends on $\delta_c(z)/\sigma_M$. The truncation used  is a good approximation of the true PDF if $\delta_c(z)/\sigma_M$ is small (and non-Gaussianity is small) but for rare events (the tails of the distribution) $\delta_c(z)/\sigma_M$ is large. One thus expect this approximation to break down for large masses, high redshift and high $\fnl$. 

\begin{figure*}[t]
\centerline{\includegraphics[width=0.5\linewidth]{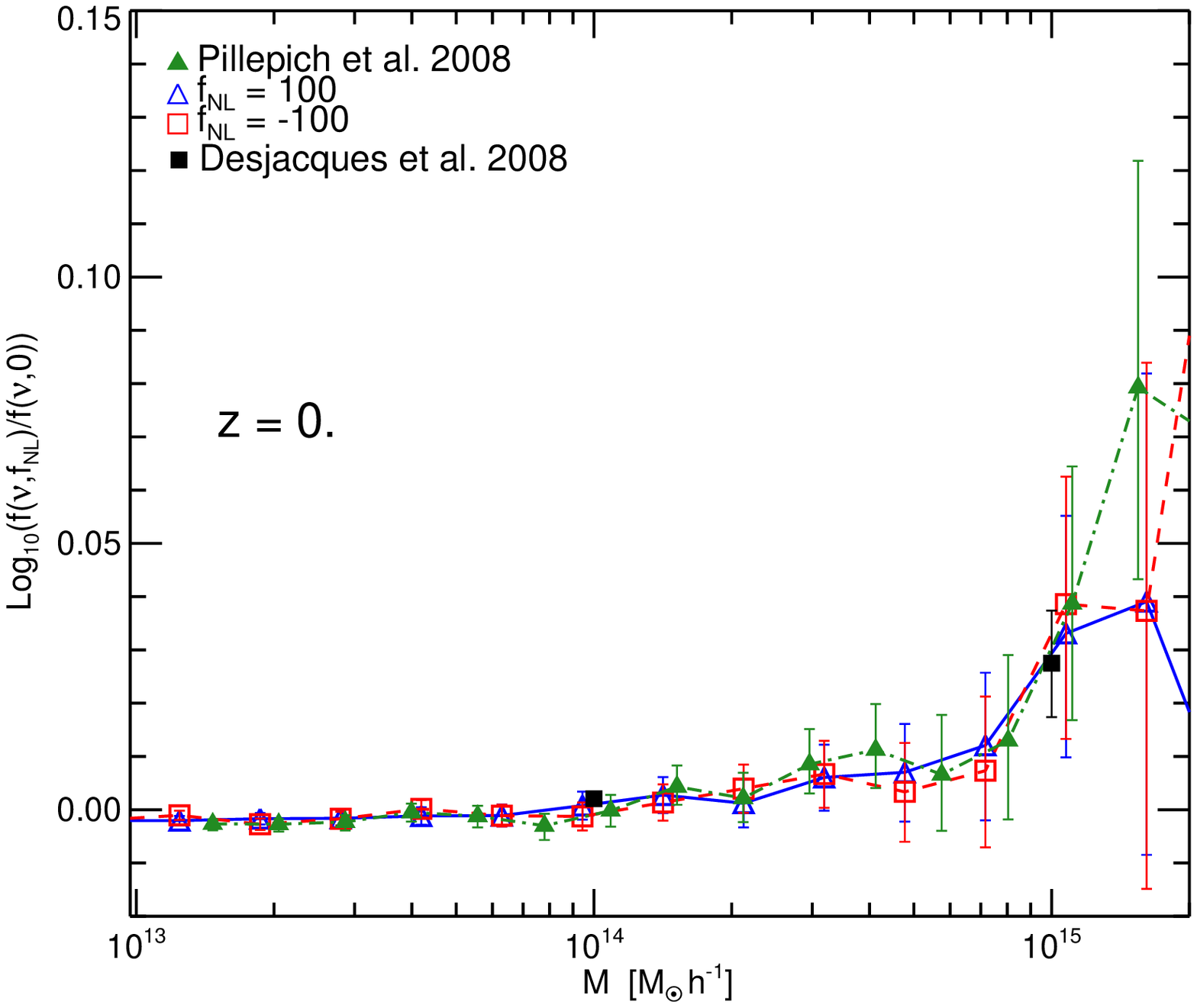}\includegraphics[width=0.5\linewidth] {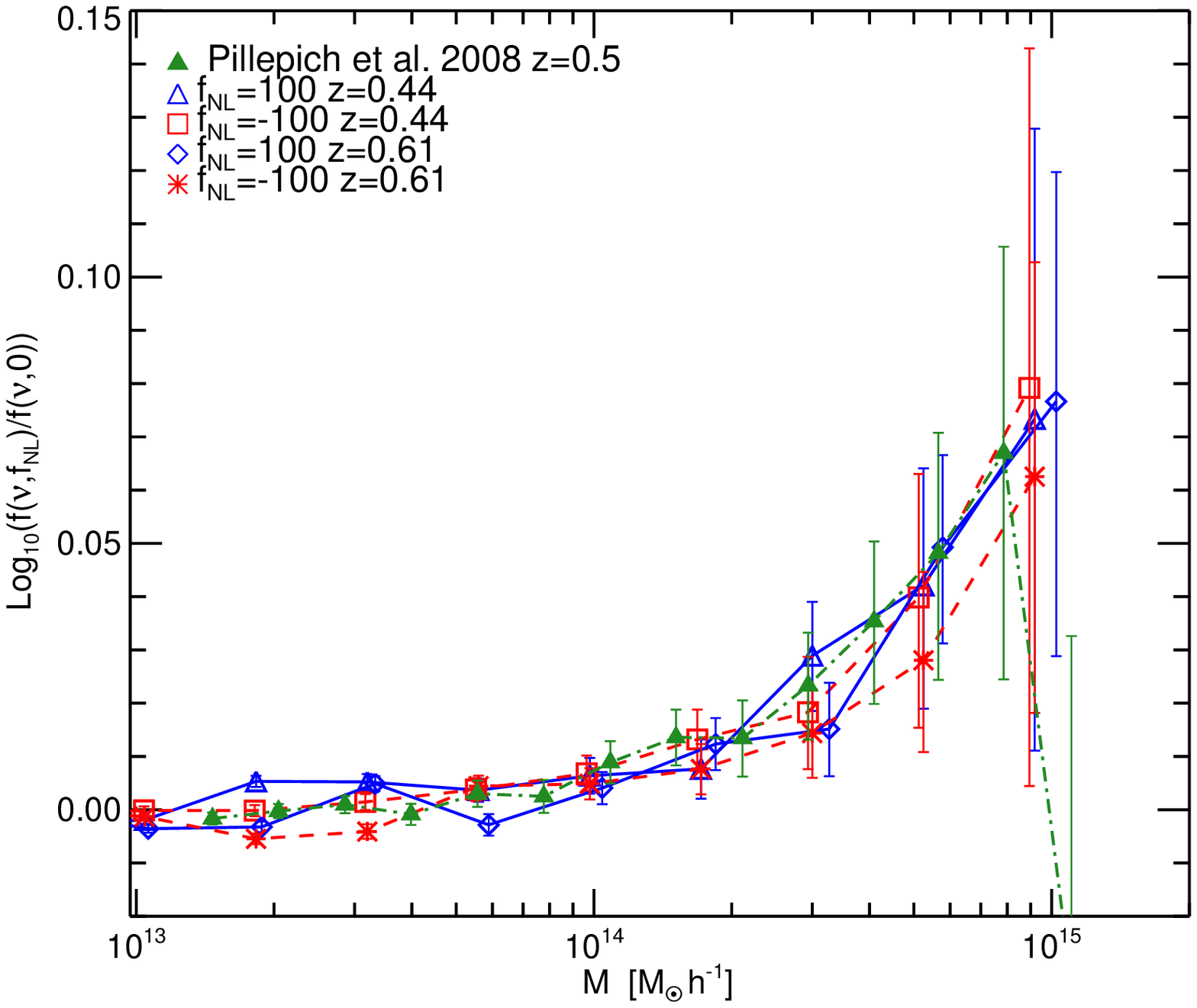}}
\caption{Correction to the Gaussian mass function as measured in different non-Gaussian simulations. There is now agreement between different simulations.  The y axis should be interpreted as $Log_{10} {\cal R}(M,z,\fnl)$. Reproduced from fig. 4 and 5 of  Ref. \cite{Grossietal09}.}
\label{fig:massfncomp}
\end{figure*}

Ref. \cite{Loverdeetal07}   quantified the range of validity of their approximation by assuming that when terms proportional to $S_3^2$ become important is no longer valid to neglect terms proportional to higher-order cumulants. Then they define the validity regime of their mass function  to be where corrections from the $S_3^2$  are unimportant. They find, as expected,  that  for very massive objects the approximation breaks down and that  the upper mass  limit for applicability of the   mass function decreases with redshift and $\fnl$. But for low masses, redshifts and $\fnl$ their formula is better than the MVJ.
On the other hand the MVJ range of validity extends to higher masses, redshifts and $\fnl$  values, as  expected, as MVJ applied the saddle point approximation to ${\cal P}(>\delta_c|M,z)$ which is an increasingly good approximation for rare objects.

Of course, the natural observable to apply this method to are not only galaxy surveys (and the clusters found there), but, especially suited, are the mass-selected large clusters surveys offered by on-going Sunyaev-Zeldovich  experiments (e.g., Planck, ACT, SPT). 

A detailed comparison with N-body simulations is the next logical step to pursue.

\subsection{Comparison with N-body simulations}

Before we proceed we should  consider that the Press \& Shechter formulation of the mass function even in the Gaussian initial conditions case, can be  significantly improved see e.g., \cite{ST99, JenkinsMF, ReedMF}. Much improved expressions have been extensively calibrated on Gaussian initial conditions N-body simulations. 
The major limitations in both the MVJ and LMSV derivations (since they follow the classic Press \& Shechter formulation) are the assumption of spherical collapse and the sharp $k$-space filtering. In addition, the excursion set improvement on the original Press \& Schechter swindle  relies  on the random-phase hypothesis, which is not satisfied for non-Gaussian initial conditions.
Since these improvements of the mass function  have not yet been generalized to generic non-Gaussian initial conditions (but work is on-going, see other contributions in this issue) the analytical results above should be used to model  fractional {\it corrections} to the Gaussian case. 

Thus the non-Gaussian  mass function, $n_{\rm NG}(M,z)$ can be written as a function of a Gaussian one, $n_G(M,z)$ (accurately calibrated on N-body simulations) with a non-Gaussian {\it correction} factor ${\cal R}$ (see e.g., \cite{VJKM01,Loverdeetal07}):
\begin{equation}
n_{\rm NG}(M,z)=n_G(M,z) {\cal R}(S_3,M,z)
\end{equation}
where
\begin{equation}
{\cal R}(S_3,M,z)=\frac{\hat{n}(M,z,\fnl)}{\hat{n}(M,z,\fnl=0)}
\end{equation}
and $\hat{n}$ is given by the MVJ or LMSV approximation. The correction ${\cal R}$ can then be calibrated on N-body simulations.

\begin{figure}[t]
\centerline{\includegraphics[width=1 \linewidth]{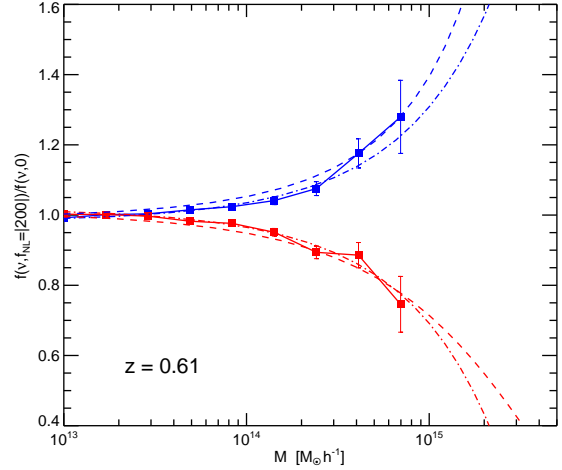}}
\caption{The points show the non-Gaussian correction to the mass function as measured in the  N-body simulations of \cite{Grossietal09}. Blue   corresponds to  $\fnl=200$ and red to  $\fnl=-200$. the dashed lines correspond to the MVJ formulation and the dot-dashed lined to the LMSV formulation. In both cases the substitution $\delta_c\longrightarrow \delta_{ec}$ has been performed.    The y axis should be interpreted as ${\cal R}(M,z,\fnl=200)$. Reproduced from fig. 7 Ref. \cite{Grossietal09}.}
\label{fig:MVJLoVsims}
\end{figure}

Ref. \cite{Grossietal09}  argue that the same correction that in the Gaussian case modifies the collapse threshold, $\delta_c$, to improve over the original Press \& Shechter formulation, may apply to the non-Gaussian correction. The  detailed  physical interpretation of this is still matter of debate in the literature \cite{LeeShandarin,Robertson09,MaggioreRiotto}. 
In summary, Ref. \cite{Grossietal09} proposes to write the non-Gaussian correction factor  for the MVJ \cite{MVJ00}case as:
\begin{eqnarray}
\label{eq:ratioMVJellips}
&&{\cal R}_{NG}(M,z,f_{NL})=\exp\left[\delta_{ec}^3
\frac{S_{3,M}}{6 \sigma_M^2}\right] \times \\
& &\!\!\!\! \left| \frac{1}{6}
\frac{\delta_{ec}}{\sqrt{1-\frac{\delta_{ec}S_{3,M}}{3}}} 
\frac{dS_{3,M}}{d\ln \sigma_{M}}  
\!+\! \sqrt{1-\frac{\delta_{ec} S_{3,M}}{3}}\right|, 
\nonumber 
\end{eqnarray}
and for the LMSV \cite{Loverdeetal07}  case:
\begin{eqnarray}
\label{eq:ratioLoVellips}
&&{\cal R}_{NG}(M,z,f_{NL})=1+\frac{1}{6}\frac{\sigma_M^2}{\delta_{ec}} \times \\
&&\left[S_{3,M}\left(\frac{\delta_{ec}^4}{\sigma_M^4} 
-2\frac{\delta_{ec}^2}{\sigma_M^2}-1\right)+
\frac{dS_{3,M}}{d \ln \sigma_M}
\left(\frac{\delta_{ec}^2}{\sigma_M^2}-1\right)\right] \nonumber
\end{eqnarray}
where $\delta_{ec}$ denotes the modified critical density for  collapse, 
which for high peaks is $\delta_{ec}\sim \delta_c \sqrt{q}$.  Ref. \cite{Grossietal09}   calibrated these expressions on N-body simulations to find  $q=0.75$. We anticipate here that the validity of this extrapolation (i.e. in terms of a correction to the critical collapse threshold) can be tested independently on the large-scale non-Gaussian halo bias  as described in \S \ref{sec:halobias}.
Note that, in both cases, in the limit of small non-Gaussianty the correction factors reduce to
\begin{equation}
{\cal R}=1+S_{3,M}\frac{\delta_{ec}^3}{6\sigma_M^2}\,.
\end{equation} 
Non-Gaussian mass functions have been computed from simulations and compared with different theoretical predictions in several works \cite{KNS07,Grossietal07,DDHS07,Desjacques, Pillepich}. In the past, conflicting results were reported, but the issue seems to have been settled: there is agreement among mass function measured from  different  non-Gaussian simulations performed by three different groups  as shown for example in Fig. (\ref{fig:massfncomp}).
As expected both MVJ and LMSV prescriptions for the non-Gaussian  correction to the mass function agree with the simulation results, provided  one make the substitution $\delta_c\longrightarrow\delta_{ec}$, with some tentative indication that MVJ may be better for very massive objects while LMSV  performs better for less rare events. This is shown in Fig. (\ref{fig:MVJLoVsims}) where the points represent measurements from N-body simulations presented in \cite{Grossietal09}. 

\subsection{Voids}

While galaxy clusters form at the highest overdensities of the primordial density field and probe the high density tail of the PDF, voids form in the low density regions and thus probe the low-density tail of the PDF. Most of the volume of the evolved universe is underdense, so it seems interesting to pay attention to the distribution of underdense regions. 
A void distribution function can be derived in an analogous way to the Press Schechter mass function by realizing that negative density fluctuations grow into voids \cite{KVJ09}: a critical underdensity $\delta_{\upsilon}$ is necessary for producing a void and this plays the role of the critical overdensity $\delta_c$ for producing bound objects (halos).  The more underdense a void is the more negative $\delta_{\upsilon}$ becomes. The precise value of $\delta_{\upsilon}(z)$ depends on the precise definition of a void (and may depend on the the observables used to find voids); realistic values of $\delta_{\upsilon}(z=0)$ are expected to be $\gtrsim -1$. In the absence of a better prescription, here,  following \cite{KVJ09}, $\delta_{\upsilon}$ is treated as a phenomenological parameter and results are shown for a range of $\delta_{\upsilon}$ values. 
To derive the non-Gaussian void probability function one proceeds as above with the only subtlety that $\delta_{\upsilon}$ is negative and that 
${\cal P}(>\delta)=1-{\cal P}(>\delta)$ thus $|d {\cal P}(<\delta)/dM|=|d {\cal P}(>\delta)/dM|$. Thus  the void PDF as a function of $|\delta_{\upsilon}|$ can be obtained from the PDF of MVJ \cite{MVJ00} or LMSV  \cite{Loverdeetal07},  provided one keeps track of the sign of each term. For example in the LMSV approximation  the void distribution function becomes \cite{KVJ09}: 
\begin{eqnarray}
     \hat{n}(R,z,\fnl)&=& \frac{9}{2\pi^2} \sqrt{\frac{\pi}{2}}
     \frac{1}{R^4} e^{-\delta_{\upsilon}^2/2\sigma_M^2} 
     \left\{  \left| \frac{d \ln \sigma_M}{d \ln M}\right| 
     \right. \nonumber \\
     &-&  \left[   \frac{|\delta_{\upsilon}|}{\sigma_M}  \frac{S_3
     \sigma_M}{6} \left(\frac{\delta_{\upsilon}^4}{\sigma_M^4} -  2
     \frac{\delta_{\upsilon}^2}{\sigma_M^2} -1 \right) \right] \nonumber \\
     &+ & \left.  \frac{1}{6} \frac{dS_3}{dM} \sigma_M \left( \frac{
     \delta_{\upsilon}^2}{\sigma_M^2} -1 \right) \right\}.
\label{eqn:generalization}
\end{eqnarray}
where the expression is reported as a function of the smoothing radius rather than the mass, since a void  Lagrangian radius is probably easier to determine  than its mass.

Note that while a positive skewness ($\fnl>0$) boosts the number of halos at the high mass end (and slightly suppress the number of low-mass halos), it is a negative skewness that will increase the voids size distribution at  the largest voids end (and slightly decrease it for small void sizes). 
Ref. \cite{KVJ09} concluded that the abundance of voids is sensitive to non-Gaussianity:  $|\delta_{\upsilon}|$ is expected to be smaller than $\delta_c$ by a factor 2 to 3. If voids probe the same scales as halos then they should provide constraints on $\fnl$ $2$ to $3$ times worse. However voids may probe slightly larger scales than halos: in many non-Gaussian models, $S_3^{\fnl=1}$ increases  with scales (see e.g., Fig. \ref{fig:ske}), compensating for the threshold.

The approach reviewed here provides a rough estimate of the fractional change in abundance due to primordial non-Gaussianity  but will not provide reliably the abundance itself. It is important to stress here that rigorously quantitative results will need to be calibrated on cosmological simulations and mock survey catalogs. 

\section{Effects on the Halo power spectrum}
\label{sec:halobias}

Recently, Refs. \cite{DDHS07, MV08} showed that primordial non-Gaussianity affects the clustering
 of dark matter halos (i.e., density extrema) inducing a scale-dependent bias for halos on large scales.
This can be seen  for example by considering halos as regions
where the (smoothed) linear density field exceeds a suitable threshold.
All correlations and peaks considered in the section
are those of the {\it initial} density field (linearly extrapolated
to the present time).
Thus  for example in the Gaussian case
\cite{Kaiser84, PolitzerWise84, JensenSzalay86}
 for high peaks we would have the following relation between the correlation function of halos of mass $M$, $\xi_{h,M}(r)$ and that of the dark matter distribution smoothed on scale $R$, corresponding to mass $M$, $\xi_R(r)$:
 \begin{equation}
\xi_{h,M}(r)\simeq b_L^2 \xi_R(r)
\end{equation}
where $b_L \simeq \delta_c/\sigma_R^2$ denotes the Lagrangian bias, although more refined expressions  can be found in e.g., \cite{MoWhite96} and  \cite{CLMP98}. 

The Lagrangian bias appears here because correlations and peaks are those of the initial density field (lineary extrapolated). Making the standard assumptions that halos move coherently
with the underlying dark matter, one can obtain the
final Eulerian bias as $b_E=1+b_L$, using the
techniques outlined in  \cite{Eetal88}, \cite{CK89}, \cite{MoWhite96} and  \cite{CLMP98}.

The two-point correlation function of regions above a high threshold has
been obtained,  for the general non-Gaussian case, in \cite{GW86}, \cite{MLB86} and \cite
{LMV88}:
\begin{equation}
\xi_{h, M}(|{\bf x}_1-{\bf x}_2|)=-1+\exp[X]
\label{eq:X}
\end{equation}
where
\begin{equation}
X=\sum_{N=2}^{\infty}
\sum_{j=1}^{N-1}\frac{\nu^N\sigma_R^{-N}}{j!(N-1)!}\xi^{(N)}
\left[^{{\bf x}_1,...,{\bf x}_1, \,\,\,{\bf x}_2,........, {\bf x}_2}_{j \,times
\,\,\,\,\,\, (N-j)\, times}\right] \;,
\end{equation}
where $\nu=\delta_c\sigma_R$.
For large separations the exponential can be expanded to first order.
This is what we will do in what follows but we will comment on this choice below.

For small non-Gaussianities\footnote{Effectively that is for  values of $f_{\rm NL}$ consistent with  observations}, 
we can keep terms up to the three-point correlation function $\xi^{(3)}$, 
obtaining that the correction to the halo correlation function, $\Delta \xi_{h}$ 
due to a non-zero three-point function is given by:
$$
\Delta \xi_{h}=\frac{\nu_R^3}{2\sigma_R^3}
\left[\xi_R^{(3)}({\bf x}_1,{\bf x}_2,{\bf x}_2)+\xi_R^{(3)}({\bf x}_1,{\bf x}_1,{\bf x}_2)
\right] 
$$
\begin{equation}
\label{eq:correxi}
 = \frac{\nu_R^3}{\sigma_R^3}\xi_R^{(3)}({\bf x}_1,{\bf x}_1,{\bf x}_2)
\end{equation}
For a general bispectrum $B(k_1, k_2, k_3)$ this yields  a correction to the power spectrum (see \cite{MV08} for steps in the derivation):
\begin{eqnarray}
\!\!\!\frac{\Delta P}{P}\!&\!=\!&\! \frac{\delta_c(z)}{{\cal M}_R(k)}\frac{1}{4 \pi^2 \sigma_R^2}\int \! dk_1 k_1^2 {\cal M}_R(k_1)  \times  \nonumber  \\
&& \int_{-1}^1\!d\mu {\cal M}_R\left(\sqrt{\alpha}\right)\frac{B_{\phi}(k_1,\sqrt{\alpha},k)}{P_{\phi}(k)}
\label{eq:fnlk}
\end{eqnarray}
where we have made the substitution $\alpha=k_1^2+k^2+2k_1k\mu$.
Here ${\cal M}_R=W_R{\cal M}$.
The  effect on the halo bias  is
$ \Delta b^L_h/b_h^L=\frac{1}{2} \frac{\Delta P}{P}$ and thus
\begin{equation}
b_{\rm h} ^{f_{\rm NL}}= 1+\frac{\Delta_c(z)}{\sigma_R^2 D^2(z)} \left[ 1+\delta_c(z)\beta_R(k) \right] \;,
\label{eq:db}
\end{equation}
where  the expression for $\beta$ can be obtained by comparing to Eq. (\ref{eq:fnlk}). The term $\Delta_c(z)/[\sigma_R^2 D^2(z)]\simeq  b^G-1 $ can be recognized as the Gaussian Lagrangian halo  bias. 

So far the derivation  is generic for all types of non-Gaussianity specified by a given bispectrum.
We can then consider specific  cases. 
 In particular for  local non-Gaussianity we obtain:
\begin{eqnarray}
\beta_R(k)&\!\!\!=\!\!\!&\frac{2 \fnl}{8\pi^2\sigma_R^2{\cal M}_R(k)}\int dk_1 k_1^2 {\cal M}_R(k_1)
P_{\phi}(k_1) \times  \nonumber \\
&&\!\!\!\!\!\!\!\!\int_{-1}^1\!d\mu
{\cal M}_R\left(\sqrt{\alpha}\right)\left[ \!\frac{P_{\phi}\left(\sqrt{\alpha}\right)}{P_{\phi}(k)} 
+ 2 \right] . 
\end{eqnarray}

\begin{figure}[t]
\centerline{\includegraphics[width=1.0\linewidth]{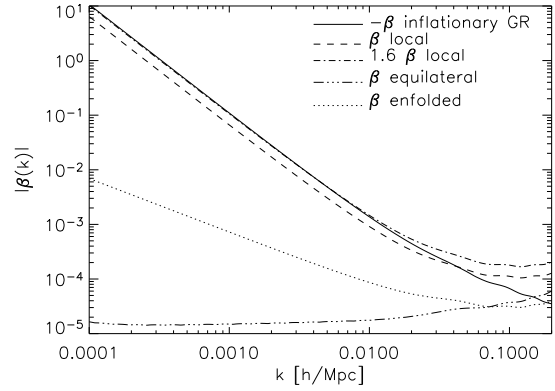}}
\caption{The scale-dependence of the large-scale halo bias induced by a non-zero bispectrum for different types of non-Gaussianity. The dashed line corresponds to the local type and the dot-dot-dot-dashed to equilateral type. Figure reproduced from \cite{VM09}.}
\label{fig:betak}
\end{figure}

Thus $\Delta b_h/b_h$ is $2 f_{\rm NL}$ times a redshift-dependent factor $\Delta_c(z)/D(z)=\delta_c(z)$,  times a $k$-- and mass--dependent  factor. The function   $\beta_R(k)$ is  shown as the dashed line  in Fig.~(\ref{fig:betak}). 
This result  for the local non-Gaussianity has been derived in at least three other ways.
Ref. \cite{DDHS07} generalize the Kaiser \cite{Kaiser84} argument of high peak bias for the local non-Gaussianity. Starting from $\nabla^2 \Phi=\nabla^2\phi +2 f_{\rm NL}\left[\phi \nabla^2 \phi +|\nabla \phi|^2\right]$ where near peaks $|\nabla \phi|^2$ is negligible they  obtain $\delta=\delta^{\fnl=0}\left[1 + 2 f_{\rm NL} \phi \right]$. The Poisson equation to convert $\phi$ in $\delta$ then gives the scale-dependence. More details  are presented elsewhere in this issue.

Ref. \cite{Slosaretal08} work in the peak-background split. This approach is especially useful to understand that it is the coupling between very large and small scales introduced by local (squeezed-cofigurations) non-Gaussianity to boost (or suppress) the peaks clustering.
In this approach, the density field can be written as $\rho(\vec{x})=\bar{\rho}(1+\delta_l+\delta_s)$ where $\delta_l$ denotes long wavelength  fluctuations and $\delta_s$ short wavelength fluctuations. $\delta_l$ is the one responsible for modulating halo formation (i.e. to boost peaks above the threshold for collapse), so the halo number density is $n=\bar{n}(1+b_L\delta_l)$ and $b_L=\bar{n}^{-1} \partial n/\partial \delta_l$.

In the local  non-Gaussian case they decompose the Gaussian field $\phi$ as  a combination of long and short wavelength fluctuations $\phi=\phi_l+\phi_s$ thus
$\Phi=\phi_l+f_{\rm NL}\phi_l^2+(1+2\fnl \phi_l)\phi_s+\fnl \phi_s^2+ const.$.
Also in this non-Gaussian case one can  split the density field $\delta$ in  $\delta_l$ and $\delta_s$  and relate this to $\fnl$ (it is easier to work in Fourier space):
$\delta_l(k)=\alpha(k)\phi_l(k)$ and $\delta_s=\alpha(k)[(1+2 \fnl \phi_l)\phi_s+\fnl \phi^2]\equiv \alpha(k)[X_1\phi_s+X_2\phi_s^2]$ the last equality giving the definition of $X_1$ and $X_2$. Note that $\delta_s$ cannot be ignored here because $\phi_l$ enters in $X_1$, in other words, local non-Gaussianity couples long and short wavelength modes. 
The local halo number density is now function of $\delta_l$, $X_1$ and  $X_2$ yielding the following result for the Lagrangian halo bias:
\begin{equation}
b_L=\bar{n}^{-1}\left[\frac{\partial n}{\partial \delta_l} +2 \fnl \frac{{\rm d} \phi_l}{{\rm d}\delta_l}\frac{\partial n}{\partial X_1}\right]= b_L^{\rm Gaussian}+
\end{equation}
$$
2\fnl \frac{{\rm d} \phi_l}{{\rm d}\delta_l}\frac{\partial \ln n}{\partial \ln \sigma_8}\equiv b_L^{\rm Gaussian}(1+2 \fnl \alpha(k)\delta_c)
$$
where $\alpha(k)$ encloses the scale-dependence of the effect.
Ref. \cite{AfshordiTolley08} rederives the ellipsoidal collapse for small deviations from Gaussianity of the local type. They find that a non-zero $\fnl$ modifies the threshold for collapse, the modification is proportional to $\fnl$. This should sound familiar from \S 3. They then use the definition $b_L={\bar n}^{-1}\partial n/\partial \delta_c$ keeping track of the fact that $\delta$ is ``modulated" by $\fnl$. 

 %
 \begin{figure*}[t]
\centerline{\includegraphics[width=0.48\linewidth]{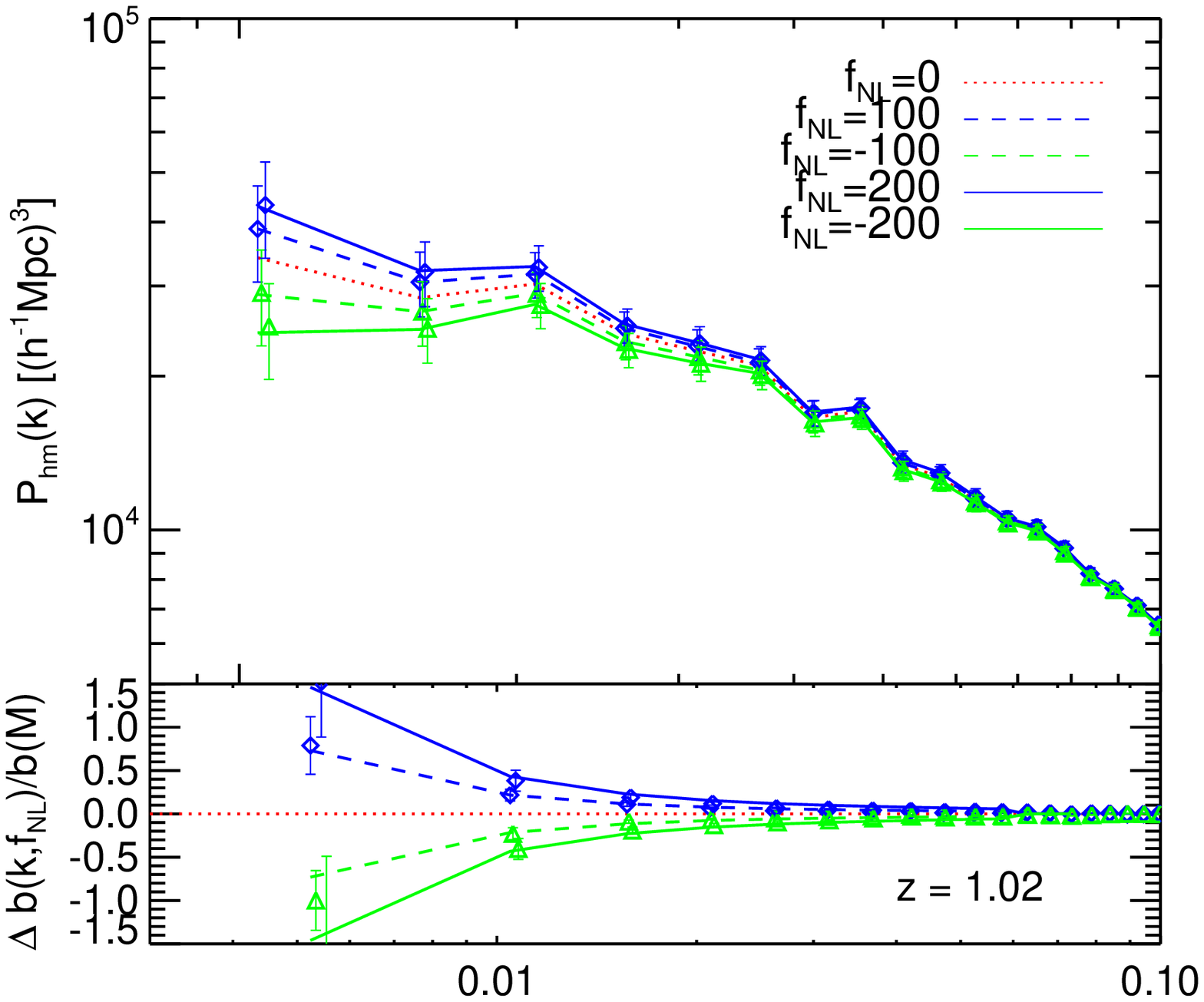}\includegraphics[width=0.61\linewidth]{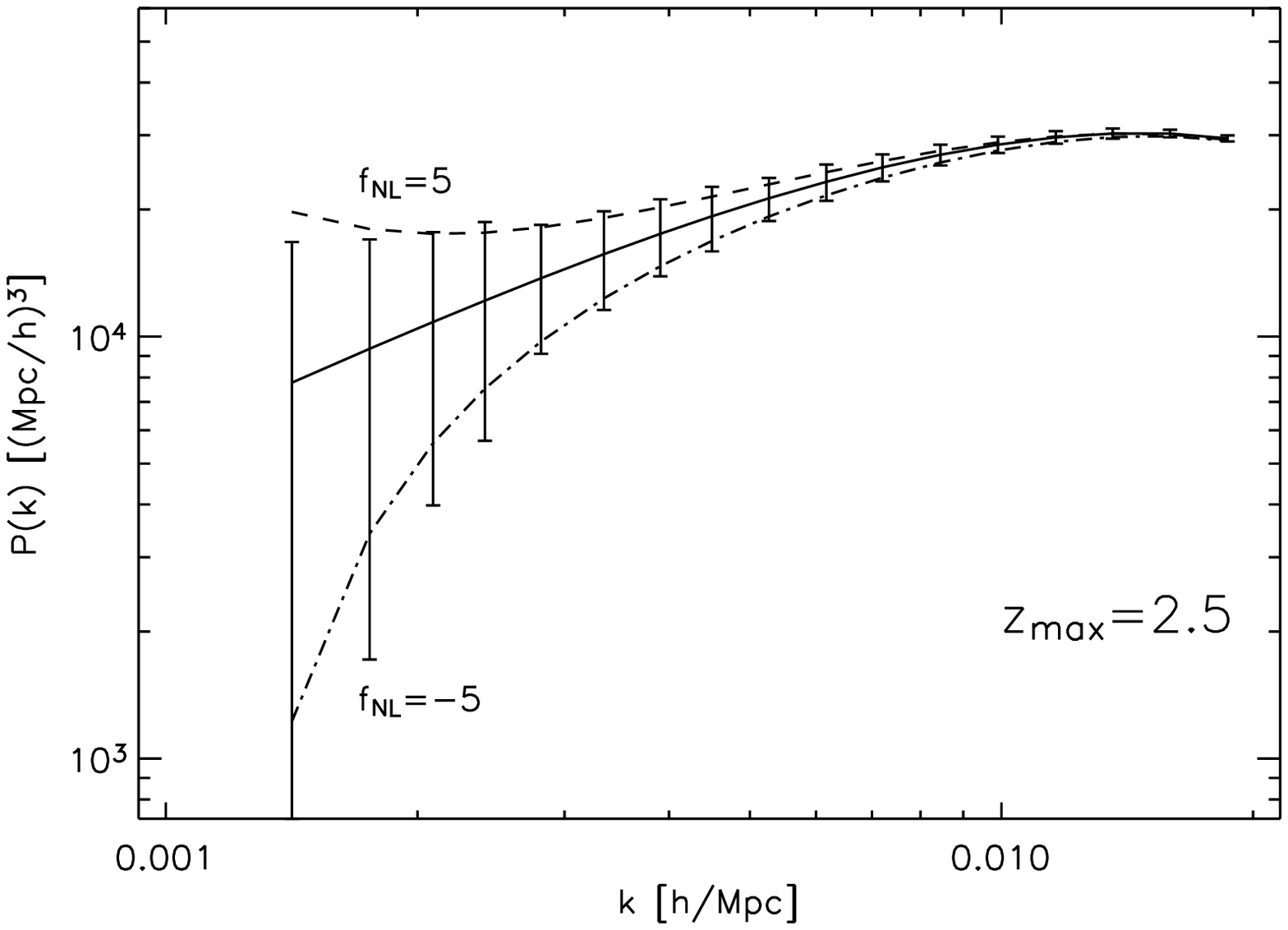}}
\caption{Effect of the non-Gaussian  halo bias  on the power spectrum. In the left-top panel we show the  halo-matter cross-power spectrum for masses above $10^{13}$ M$_{\odot}$ at $z=1.02$. The left-bottom panel shows the ratio of the non-Gaussian to Gaussian bias. Figure reproduced from \cite{Grossietal09}. The $\fnl$ values reported in the figure legend should be interpreted as $\fnl^{LSS}$. On the right panel we show the expected effect and error-bars for the large-scale power spectrum for a survey like LSST. Figure reproduced from \cite{LSSTBook}.}
\label{fig:pkNG}
\end{figure*}
The effect of the non-Gaussian halo bias on the power spectrum is shown in Fig.(\ref{fig:pkNG}) where the points are measurements from an N-body simulations  of \cite{Grossietal09} (see  figure caption for more details).

The above result 
\begin{equation}
\Delta b=  \fnl \delta_c (b_G^{\rm Gaussian} -1)\beta_R^{\fnl=1}(k)
\label{eq:NGHBias0}
\end{equation}
where $\beta_R(k)=\fnl \beta_R^{\fnl=1}(k)$,
can be improved in several ways.

First of all, we have not made any distinction between the redshift at which the object is being observed ($z_o$) and that at which is being formed ($z_f$). Except for the rarest events this should be accounted for.  The Gaussian Lagrangian bias expression used so far is an approximation, a more accurate expression is \cite{Eetal88, CK89, MoWhite96}
\begin{equation}
b^G_{L,h}(z_o,M,z_f) = \frac{1}{D(z_o)}\left[\frac{\delta_c(z_f)}{\sigma_M^2}
-\frac{1}{\delta_c(z_f)}\right] \;. 
\label{eq:gaussian_halo_bias_full}
\end{equation}

Then, the halo bias expressions are  derived within the  ``classical" Press \& Schechter theory, as we have seen in \S 3, subsequent improvements on the mass function can be seen as a correction to the collapse threshold. In the expression for the Gaussian  halo bias  
$b_L^{G}={\bar n}^{-1}\partial n^{G}/\partial \delta_c$ one can consider mass functions that are better fit to simulations than the standard Press \& Schecter one obtaining:
$$
 b^G_{L,h}(z_o,M,z_f)=\frac{1}{D(z_o)}
\left[\frac{q\delta_c(z_f)}{\sigma_M^2}-\frac{1}{\delta_c(z_f)}\right]  
$$
\begin{equation}
+ \frac{2p}{\delta_c(z_f)D(z_o)} \left[1 + 
\left(\frac{q\delta_c^2(z_f)}{\sigma_M^2} 
\right)^p\right]^{-1} \;. 
\end{equation}
The parameters $q$ and $p$ account for non-spherical 
collapse and fit to numerical simulations yield $q\sim 0.75$, $p=0.3$ 
e.g., \cite{ST99}.  
In this expression the term in the second line is usually sub-dominant. The term ``$-1/\delta_c$" in  the first line is known as ``anti-bias", and it becomes  negligible for  old halos $z_f \ll z_o$. 
Note that  by including the anti-bias correction in $b_g$ of Eq. \ref{eq:NGHBias0} one recovers the ``recent mergers" approximation of Ref. \cite{Slosaretal08}.

\begin{figure*}[t]
\centerline{\includegraphics[width=0.4\linewidth]{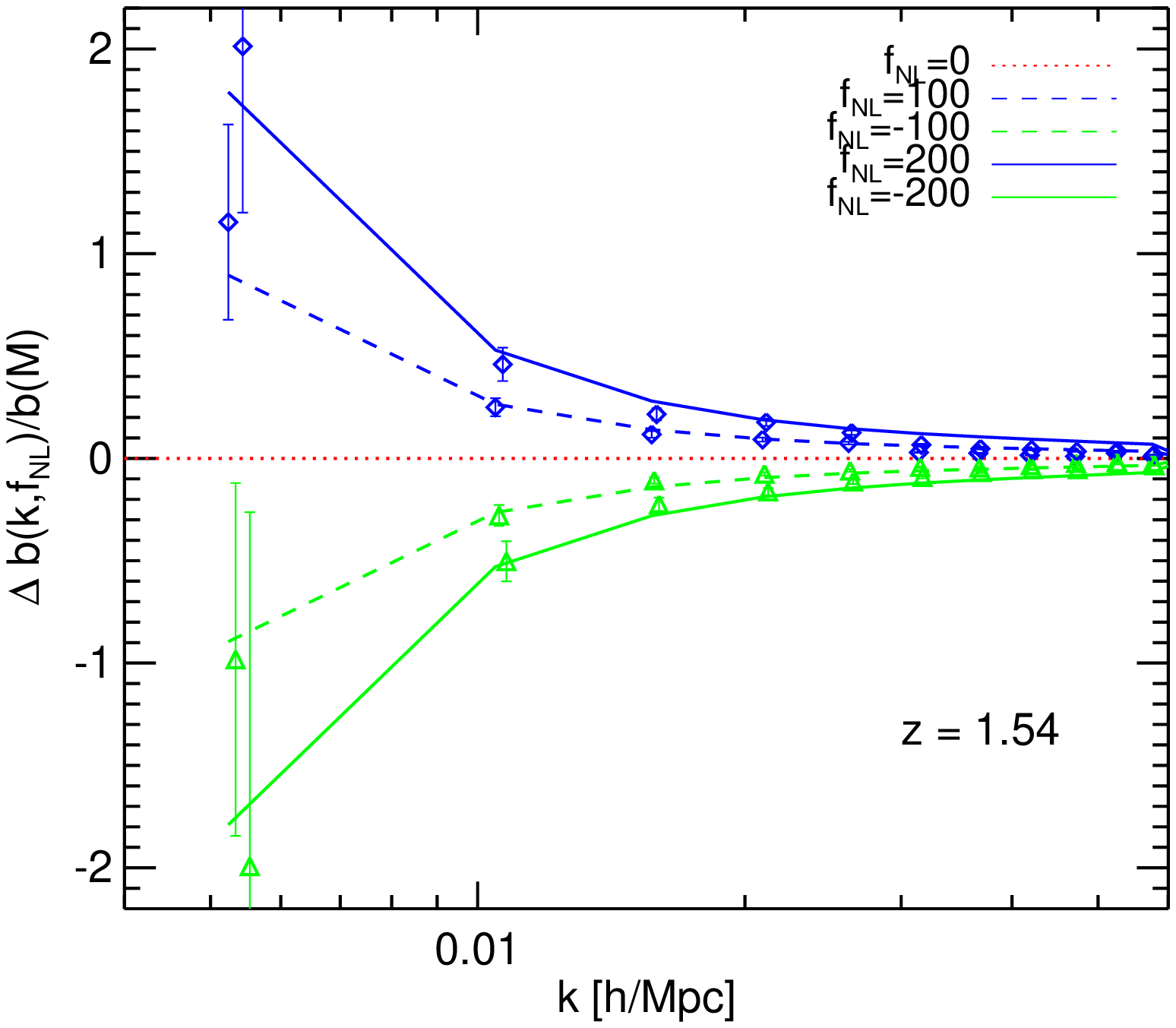}\includegraphics[width=0.4\linewidth]{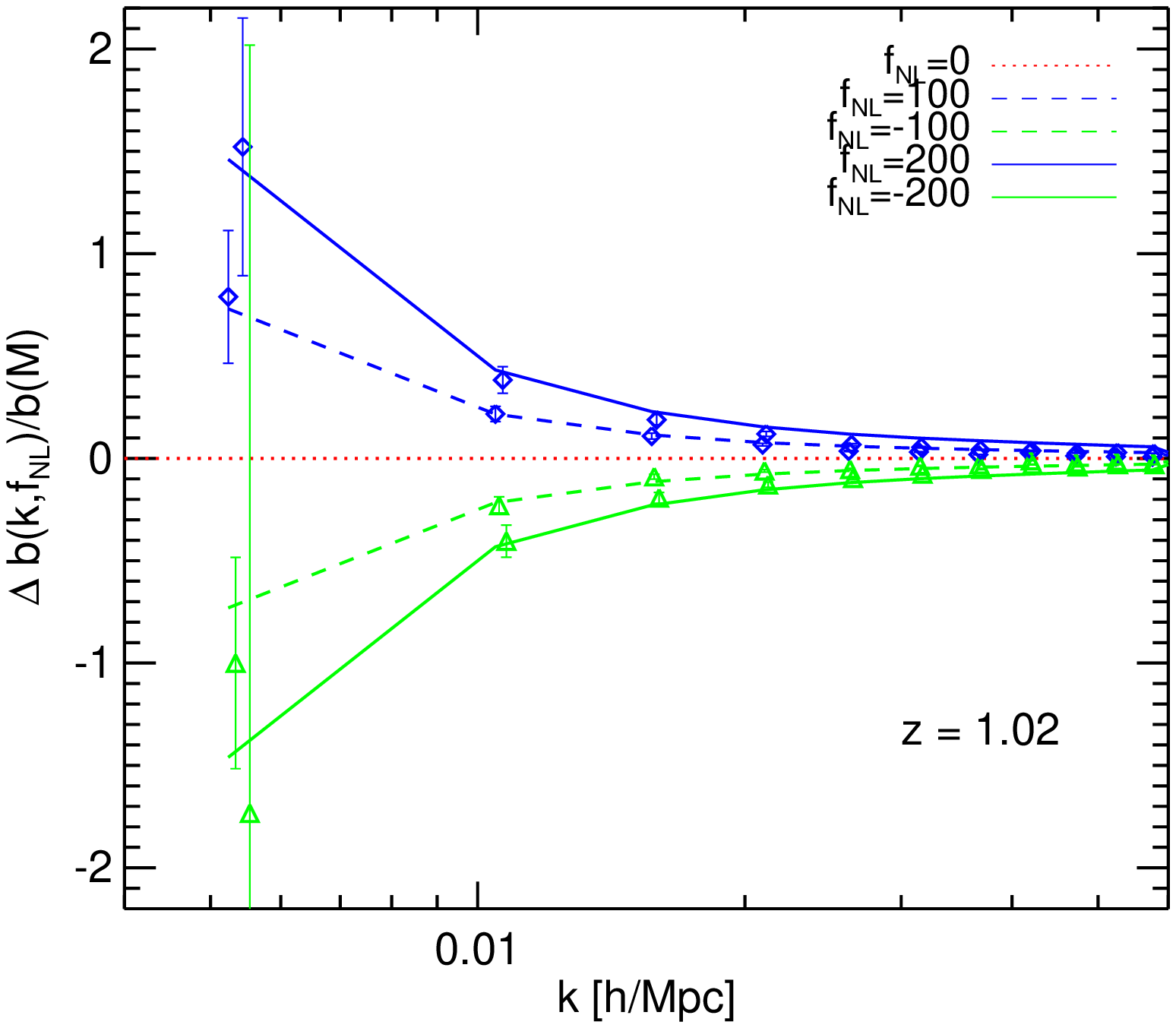}\includegraphics[width=0.4\linewidth]{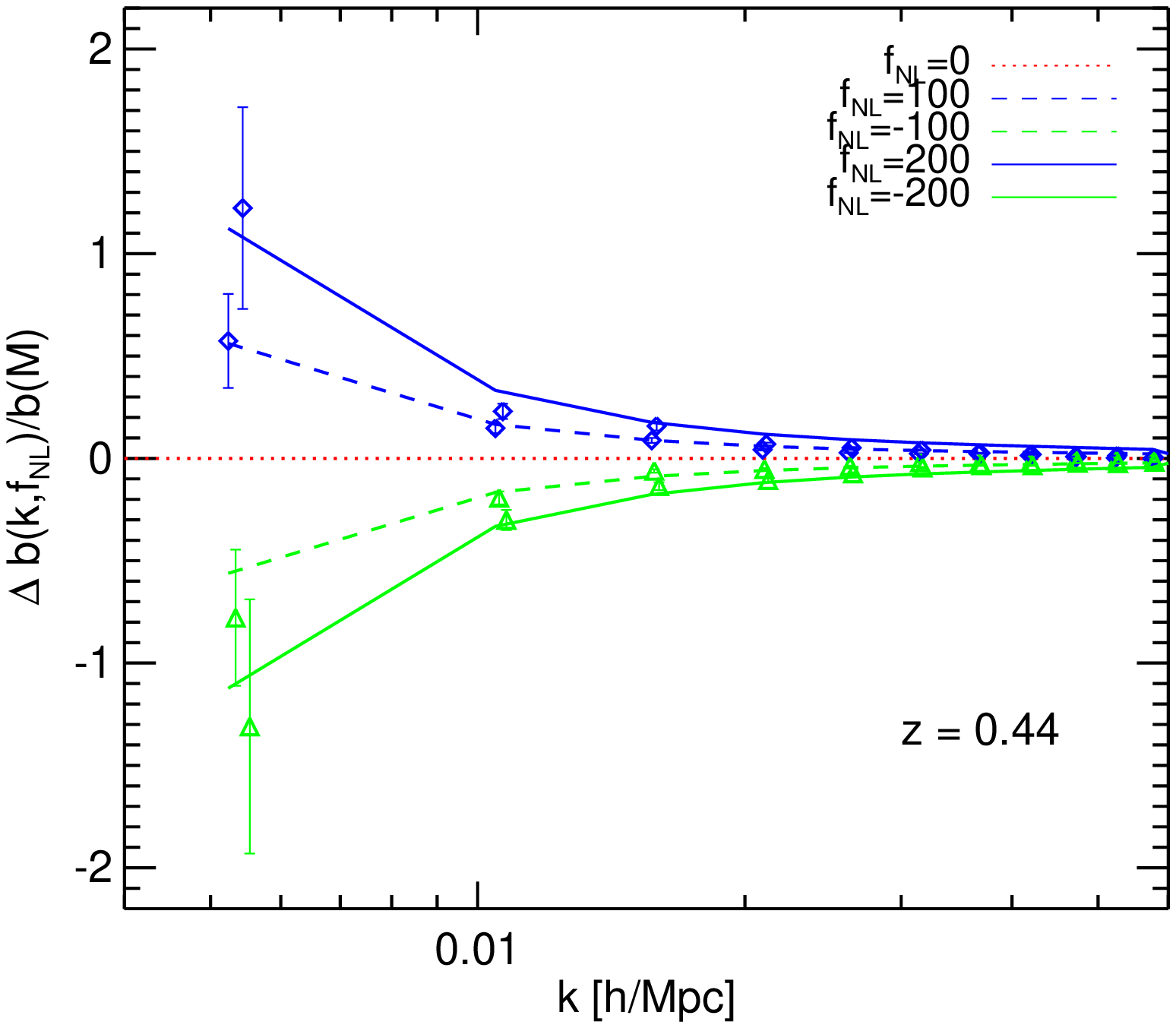}}
\caption{The quantity $\Delta b/b$ as function of $k$, for simulation snapshots at $z=0.44$, $1.02$ and $1.54$. Simulation outputs and theory lines are shown for $f_{NL}=\pm100$ and $f_{NL}=\pm  200$. Figure reproduced from \cite{Grossietal09}. The $\fnl$ values reported in this figure legend should be interpreted as $\fnl^{LSS}$.}
\label{fig:compsimbias}
\end{figure*}

The  same correction should also apply to the non-Gaussian
correction to the halo bias:
\begin{equation}
\Delta b=  f_{\rm NL} 
q'\delta_c (b^G_{h} - 1 )\beta_R^{\fnl=1}(k)\,.
\end{equation} 
where $q'$  should coincide with $q$ above; it  can be calibrated to N-body simulations and is found indeed  to be $q' = 0.75$ \cite{Grossietal09}. The non-Gaussian halo bias prediction and results from  N-body simulations  with local non-Gaussianity  are shown in Fig. (\ref{fig:compsimbias}).

Finally one may note that for $f_{\rm NL}$ large and negative, 
Eq.~(\ref{eq:fnlk}, \ref{eq:db}) would formally yield $b_h^{f_{\rm NL}}$ and 
$P_h(k)$ negative on large enough scales. 
This is a manifestation of the breakdown of the approximations made: 
a) all correlations of higher order than the bispectrum were neglected: 
for large NG this truncation may not hold; b) The exponential in 
Eq.~(\ref{eq:X}) was expanded to linear order. 
This however could be easily corrected for, remembering that the $P(k)$ 
obtained from Eq.~(\ref{eq:fnlk}) is in reality the Fourier transform of 
$X$, the  argument of the exponential. 
One would then compute the halo correlation function using Eq.~(\ref{eq:X}) 
and Fourier transforming back to obtain the halo power-spectrum.

So far we have concentrated on local non-Gaussianity, but the expression of Eqs. (\ref{eq:fnlk}, \ref{eq:db}) is more general. Using this formulation, 
Ref. \cite{VM09} computed the quantity $\beta_R^{\fnl=1}(k)$ for several types of non-Gaussianity (equilateral, local and enfolded); this is shown in Fig. (\ref{fig:betak}). It is clear that the non-Gaussian halo-bias effect has some sensitivity to the bispectrum shape, for example the effect for the  equilateral type of non-Gaussianity  is suppressed by orders of magnitude compared to the local-type and the flattened case is somewhere in the middle.  Fig. (\ref{fig:betak}) also shows a type of non-Gaussianity  arising from General-relativistic (RG) corrections on scales comparable to the Hubble radius. Note that perturbations on super-Hubble scales are initially needed in order to ``feed" the GR correction terms. In this respect the significance of this contribution is analogous to the well-known large-scale anti-correlation between CMB temperature and E-mode polarization: it is a consequence of the properties of the inflationary mechanism to lay down the primordial perturbations.   This effect has the same magnitude as a local non-Gaussianity with $\fnl \gtrsim 1$.

The next logical step is then to ask how well present or forthcoming data  could constrain non-Gaussianity using the halo-bias effect. It is interesting to note  that surveys that aim at measuring Baryon Acoustic Oscillations (BAO) in the galaxy distribution to constrain dark energy are well suited to also probe non-Gaussianity: they cover large volumes  and their galaxy number density is well suited so that on the scale of interest (both for BAO and non-Gaussianity) shot noise does not dominate the signal.
Photometric surveys are also well suited: as the non-Gaussian signal is localized at very large scales and is a smooth function of $k$, the photo-z smearing effects are unimportant.

The theory developed so far describes the clustering of halos while we observe galaxies. Different galaxy populations occupy dark matter halos following different prescriptions. If we think in the halo-model framework (e.g., \cite{CooraySheth02} and references therein) at very large scales only the ``two halo'' contribution matters and the details of the halo occupation distribution (the ``one-halo'' term) become unimportant.

\begin{figure}[t]
\centerline{\includegraphics[width=1.0\linewidth]{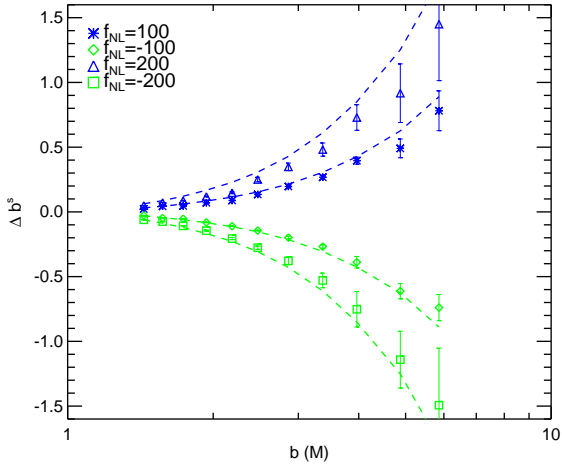}}
\caption{Non-Gaussian halo bias correction as function of the gaussian halo bias. Figure reproduced from \cite{Grossietal09}.The $\fnl$ values reported in the figure legend should be interpreted as $\fnl^{LSS}$.}
\label{fig:dbvsb}
\end{figure}

What is important to keep in mind is that the effect of  the non-Gaussianity parameter one wants to measure, $\fnl$,  is fully degenerate with the value of the Gaussian (small scales) halo bias. Fig. (\ref{fig:dbvsb}) shows the dependence of the non-Gaussian correction on the Gaussian bias.

Thus highly biased tracers will show a a larger non-Gaussian effect for the same  $\fnl$ value. Of course for a given cosmological model  the Gaussian bias can be measured accurately by comparing the predicted dark matter power spectrum with the observed one. Alternatively, two differently biased tracers can be used in tandem to disentangle the two effects \cite{Seljakfnlcv,Slosarfnlcv}.

Since clustering amplitude may depend on the entire halo history, it becomes then interesting to model in details the dependence of the effect on the halo merger tree (Reid et al., in preparation).

\subsection{Outlook for the future}

How well can this method do to constrain primordial non-Gaussianity compared with the other techniques presented here?
The Integrated Sachs Wolfe (ISW) effect offers a window to probe  clustering on the largest scales (where the signal is large); on the other hand, a measurement of clustering of tracers of dark matter halos   is a  very direct window into this effect.
A Fisher matrix approach \cite{CVM08,AfshordiTolley08,Slosaretal08} shows that the ISW signal  is weighted at relatively low redshift (where dark energy starts dominating) while the non-Gaussian signal grows with redshift, thus making  the shape of the halo power spectrum a more promising tool.
An overview of current constraints from different approaches  can be found in table \ref{tab:currentlocalfnl} and future forecasts in table \ref{tab:futurelocalfnl}, for non-Gaussianity of the local type.
Large, mass-selected  cluster samples as produced by SZ-based experiments will provide a optimally suited data-set for this technique (see e.g. \cite{fedeli}).  
\begin{table*}
 \caption{\label{tab:currentlocalfnl} Current recent $2-\sigma$ constraints on local $\rm fnl$}
 \begin{tabular}{ccc}
 Data/method & $\fnl$ & reference \\
 \hline
 \hline
 Photometric LRG - bias & $63^{+54+101}_{-85-331}$ & Slosar et al. 2008 \\
 Spectroscopic LRG- bias & $70^{+74 +139}_{-83 -191}$& Slosar et al. 2008\\
 QSO - bias& $8^{+26+47}_{-37-77}$& Slosar et al. 2008\\
 combined & $28^{+23+42}_{-24-57}$&Slosar et al. 2008\\
 \hline
 NVSS--ISW  &$105^{+647+755}_{-337-1157}$ &  Slosar et al. 2008\\
 NVSS--ISW & $ 236\pm 127 (2-\sigma)$&Afshordi\&Tolley 2008\\
 \hline
 WMAP3-Bispectrum & $30\pm 84$& Spergel et al. (WMAP) 2007 \\
 WMAP3-Bispectrum & $32 \pm 68$ & Creminelli et al.  2007\\
 WMAP3-Bispectrum &$87\pm 60$ & Yadav \& Wandelt  2008 \\
 WMAP-Bispectrum &$38\pm 42$& Smith et al.  2009\\
 WMAP5-Bispectrum & $51\pm 60$& Komatsu et al. (WMAP) 2008  \\
 WMAP5-Minkowski & $-57 \pm 121$ & Komatsu et al. (WMAP) 2008\\
 \end{tabular}
 \end{table*}

\begin{table*}
\caption{\label{tab:futurelocalfnl} Forecasts  $1-\sigma$ constraints on local $\fnl$}
 \begin{tabular}{ccc}
 Data/method & $\Delta \fnl$ ($1-\sigma$) & reference\\
 \hline
 \hline
 BOSS--bias & $18$& Carbone et al. 2008\\
 ADEPT/Euclid--bias &$1.5$ &  Carbone et al. 2008 \\
 PANNStarrs --bias  &$3.5$ & Carbone et al. 2008\\
 LSST--bias  &$0.7$ & Carbone et al. 2008\\
\hline
 LSST-ISW&$7$& Afshordi\& Tolley 2008\\
\hline
 BOSS--bispectrum&$35$& Sefusatti \& Komatsu 2008\\
ADEPT/Euclid --bispectrum & $3.6$ & Sefusatti \& Komatsu 2008\\
 Planck-Bispectrum &$3$ & Yadav et al . 2007  \\ 
 BPOL-Bispectrum&$2$  & Yadav et al . 2007   \\
 \end{tabular}
 \end{table*}


%

%
 

While for a given $\fnl$ model such as the local one, methods that exploit  the  non-Gaussian bias  seem to yield the smallest error-bars for large-scale structure,  it should be kept in mind that the bispectrum  can be used to investigate the full configuration dependence of $\fnl$ and thus is  a very powerful  tool to discriminate between different type of non-Gaussianity. In addition CMB-bispectrum  and halo bias test non-Gaussianity on very large scales while the large scale structure bispectrum mostly probes mildly non-linear scales. As primordial non-Gaussianity may be scale-dependent, all these techniques are highly complementary.

The above estimates assume that the underlying cosmological model is known. The large-scale shape of the power spectrum can be affected by cosmology. Carbone et al. (in prep.) explore possible degeneracies between $\fnl$ and cosmological parameters. They find that the parameters that are most strongly correlated with $\fnl$ are parameters describing dark energy clustering, neutrino mass and running of the primordial power spectrum spectral slope.  For surveys that cover a broad redshift range the  error on $\fnl$ degrade little when marginalizing over these extra parameters: the peculiar redshift dependence of the non-Gaussian signal lifts the degeneracy.
\section{Conclusions}
\label{sec:conclusions}
A natural question to ask at this point may be ``what observable will have better chances to constrain primordial non-Gaussianity?"

In principle the abundance of rare events is a very powerful probe of non-Gaussianity; however, in practice, it is limited by  the practical difficulty of determining the mass of the observed objects and its corresponding large uncertainty in the determination. This point is stressed e.g.,  in \cite{Loverdeetal07}. With the advent of high-precision measurements of  gravitational lensing  by massive clusters, the mass uncertainty, at least for small to moderate size clusters samples can be greatly reduced. Forthcoming Sunyaev-Zeldovich experiments will provide large samples of mass-selected  clusters  which could then be followed up by lensing mass measurements (see e.g., \cite{Sealfonetal06,maxbcg}). So far there is only one very high redshift ($z=1.4$) very massive $M \simeq 8 \times 10^{14} M_{\odot}$ with high-precision mass determination via  gravitational lensing \cite{Jeeetal09}.  Ref. \cite{JV09} pointed out that this object is extremely rare, for Gaussian initial conditions there should be $0.002$ such objects or less  in the surveyed area, which is uncomfortably low probability. But the cluster mass is very well determined:  a non-Gaussianity still compatible with CMB constraints could bring the probability of observing of the object  to more comfortable values.  This result should be interpreted as a ``proof of principle'' showing  that   this  a potentially powerful avenue to pursue.

The measurement of the three-point correlation function allows one to map directly the shape-dependence of the bispectrum. For large-scale structures the limiting factors are the large non-Gaussian  contribution induced by gravitational evolution and the uncertainly of the non-linear behavior of galaxy bias.

The halo-bias approach can yield  highly competitive constraints, but it is less sensitive to the bispectrum shape.   Still, the big difference in the magnitude and shape of the scale-dependent biasing factor between different non-Gaussian models implies that the halo bias can become a useful tool to study shapes when combined with e.g.  measurements of the CMB bispectrum. Table \ref{tab:halobiascmb} highlights this complementarity. For example, one could  envision different scenarios.
\begin{table}
 \caption{Forecasted non-Gaussianity constraints: A) \cite{YadavWandelt08} B) \cite{CVM08} C) \cite{CMBPol,Sefusattietal09} E) \cite{VM09}) e.g., \cite{MangilliVerde09} \label{tab:halobiascmb}}
 \begin{tabular}{l c c c c}
 \hline
   &  \multicolumn{2}{c}{CMB Bispectrum} & \multicolumn{2}{c}{Halo bias} \\ 
   \hline
   \small{type NG} & \small{Planck} & \small{BPol} & \small{Euclid} & \small{LSST} \\
   \hline \hline
& \multicolumn{4}{c}{1-$\sigma$ errors} \\
\hline
\small{\rm{Local}}& $3^{A)}$ & $2^{A)}$ & $1.5^{B)}$ & $0.7^{B)}$\\
\small{\rm{Equilateral}}$\!\!\!$ & $25^{C)}$ &$14^{C)}$& $-$ &$-$\\
\small{\rm{Enfolded}} & ${\cal O}10$ & ${\cal O} 10$& $39^{E)}$ & $18^{E)}$\\
\hline
 & \multicolumn{4}{c}{\#$\sigma$ detection} \\
\hline
\small{\rm{GR}} & N/A & N/A & $1^{E)}$ & $2^{E)}$\\
\small{\rm{secondaries}}$\!\!\!\!\!$ & $3^{F)}$ & $5^{F)}$& N/A & N/A\\
\end{tabular} 
\end{table} 

If non-Gaussianity is local with negative $f_{NL}$  and  CMB obtains a detection, then the  halo bias approach should also give a high-significance detection (GR correction and primordial contributions add up), while if it is local  but with positive $f_{NL}$, the halo-bias approach could give a lower statistical significance for small $f_{NL}$ as the GR correction contribution has the opposite sign.

If CMB detects $\fnl$ at the level of $\sim 10$ and of a form that is close to local, but halo bias does not detect it, then the CMB bispectrum is given by secondary effects.

If CMB detects non-Gaussianity but is not of the local type, then  halo bias can help discriminate between equilateral and enfolded shapes: if halo bias  sees a signal it indicates enfolded type,; if halo bias does not see a signal  it indicates equilateral type. Thus even a non-detection of the halo-bias effect, in combination with CMB constraints can have an important discriminatory power.
 
\begin{figure*}[t]
\centerline{\includegraphics[width=1.0\linewidth]{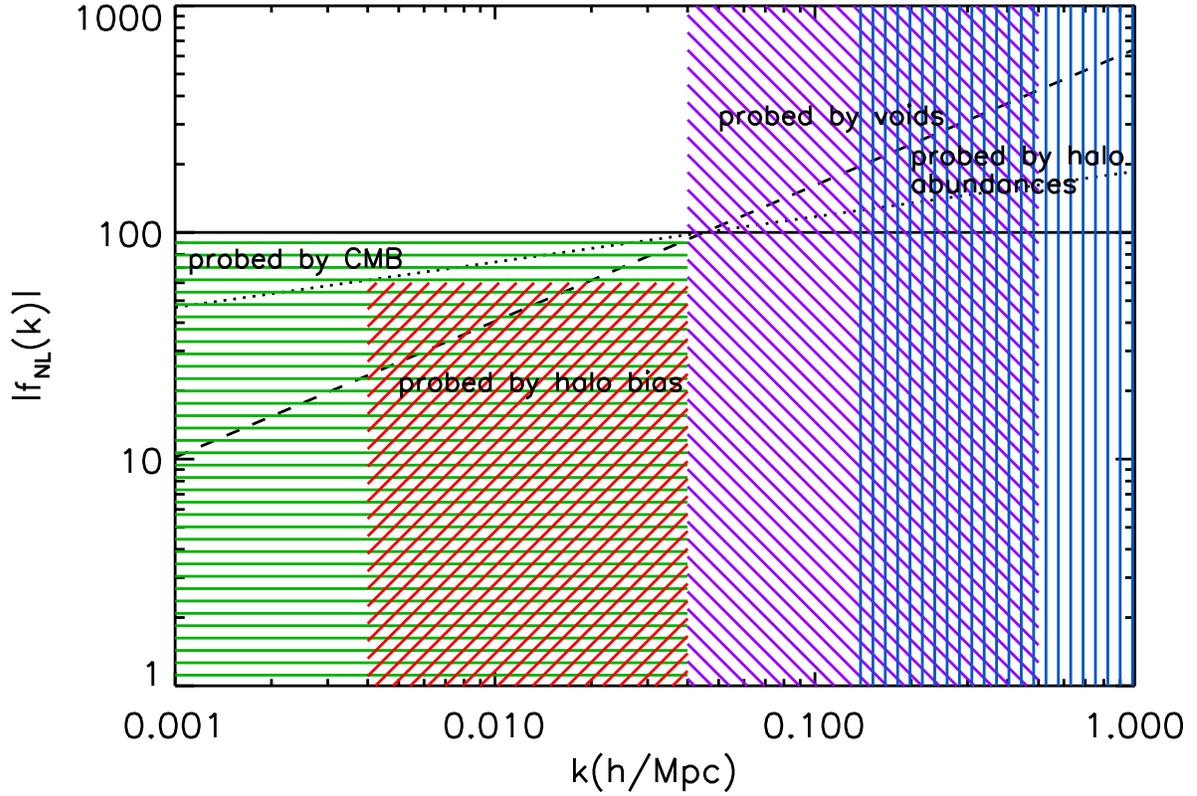}}
\caption{Scale-dependent $\fnl$ and scales probed by different approaches mentioned here. The solid line has $n_g=0$, the dotted line has $n_g=0.2$ and the dashed one has $n_g=0.6$. Hashed areas for CMB and halo-bias show allowed regions.}
\label{fig:fnlscale}
\end{figure*}

 In any case, if the simplest inflationary scenario holds,  for surveys like Euclid and LSST,   the halo-bias approach is expected to detect a non-Gaussian signal  very similar to the local type signal with an amplitude of $f_{NL}\sim -1.5$ which is  due to large-scales GR corrections  to the Poisson equation. This effect should leave no imprint in the CMB: once again the combination of the two observable can help enormously to discriminate among models for the origin of cosmological structures.

In addition we should bear in mind that 
non-Gaussianity may be scale-dependent. In fact for models like DBI inflation it is expected to be scale-dependent.  A proposed  parameterization of the scale-dependence of non-Gaussianity is given by :
\begin{equation}
B_{\phi}(\vk_1,vk_2,\vk_3)=\fnl \left(\frac{K}{k_p}\right)^{n_{ng}}F(\vk_1,\vk_2,\vk_3)
\end{equation}
where $K=(k_1 k_2 k_3)^{1/3}$ \cite{Sefusattietal09}, $k_p$ denotes the pivot and $n_{ng}$ the slope or running of non-Gaussianity ,although other authors prefer to use $K=(k_1+k_2+k_3)/3$ \cite{Loverdeetal07, AfshordiTolley08} as for squeezed configurations $K\ne 0$. It is still an open issue which parameterization is better in practice.\\

In any case different observables probe different scales (see Fig.\ref{fig:fnlscale})  and their complementary means that ``the combination is more than the sum of the parts".

What is clear, however, is that the thorny systematic effects that enter in all these approaches will require that a variety of complementary avenues be taken to establish a robust detection of primordial non-Gaussianity.\\

\noindent
{\bf Acknowledgements.} This work is supported by MICCIN grant AYA2008-03531 and FP7- 
IDEAS-Phys.LSS 240117. I would like to  thank my  closest collaborators in many of the articles  reviewed here: Carmelita Carbone, Klaus Dolag, Margherita Grossi,  Alan Heavens, Raul Jimenez,  Marc Kamionkowski, Marilena LoVerde, Sabino Matarrese,  Lauro Moscardini, Sarah Shandera and my collaborators for the reported  work-in-progress: Carmelita Carbone, Olga Mena, Beth Reid.

\end{document}